\input BoxedEPS.tex
\SetRokickiEPSFSpecial
\HideDisplacementBoxes
\documentstyle[12pt]{article}

\def \cp89{{\it CP Violation,} edited by C. Jarlskog (World Scientific,
Singapore, 1989)}

\def \f79{{\it Proceedings of the 1979 International Symposium on Lepton and
Photon Interactions at High Energies,} Fermilab, August 23-29, 1979,
ed. by T. B. W. Kirk and H. D. I. Abarbanel (Fermi National
Accelerator Laboratory, Batavia, IL, 1979}
\def \hb87{{\it Proceeding of the 1987 International Symposium on Lepton and
Photon Interactions at High Energies,} Hamburg, 1987, ed. by W. Bartel
and R. R\"uckl (Nucl. Phys. B, Proc. Suppl., vol. 3) (North-Holland,
Amsterdam, 1988)}

\def \ichep72{{\it Proceedings of the XVI International Conference on High
Energy Physics}, Chicago and Batavia, Illinois, Sept. 6 -- 13, 1972,
edited by J. D. Jackson, A. Roberts, and R. Donaldson (Fermilab,
Batavia, IL, 1972)}

\def \ite{{\it et al.}}
\def \lkl87{{\it Selected Topics in Electroweak Interactions} (Proceedings of
the Second Lake Louise Institute on New Frontiers in Particle Physics,
15 -- 21 February, 1987), edited by J. M. Cameron \ite~(World
Scientific, Singapore, 1987)}
\def \ky85{{\it Proceedings of the International Symposium on Lepton and
Photon Interactions at High Energy,} Kyoto, Aug.~19-24, 1985, edited
by M.  Konuma and K. Takahashi (Kyoto Univ., Kyoto, 1985)}

\def \np#1#2#3{Nucl. Phys. {\bf#1}, #2 (#3)}

\def \plb#1#2#3{Phys. Lett. B {\bf#1}, #2 (#3)}
\def \pr#1#2#3{Phys. Rev. {\bf#1}, #2 (#3)}
\def \prd#1#2#3{Phys. Rev. D {\bf#1}, #2 (#3)}
\def \prl#1#2#3{Phys. Rev. Lett. {\bf#1}, #2 (#3)}

\def \zpc#1#2#3{Z. Phys. C. {\bf#1}, #2 (#3)}

\def \si90{25th International Conference on High Energy Physics, Singapore,
Aug. 2-8, 1990}
\def \slc87{{\it Proceedings of the Salt Lake City Meeting} (Division of
Particles and Fields, American Physical Society, Salt Lake City, Utah,
1987), ed. by C. DeTar and J. S. Ball (World Scientific, Singapore,
1987)}
\def \slac89{{\it Proceedings of the XIVth International Symposium on
Lepton and Photon Interactions,} Stanford, California, 1989, edited by
M.  Riordan (World Scientific, Singapore, 1990)}
\def \smass82{{\it Proceedings of the 1982 DPF Summer Study on Elementary
Particle Physics and Future Facilities}, Snowmass, Colorado, edited by
R.  Donaldson, R. Gustafson, and F. Paige (World Scientific,
Singapore, 1982)}
\def \smass90{{\it Research Directions for the Decade} (Proceedings of the
1990 Summer Study on High Energy Physics, June 25--July 13, Snowmass,
Colorado), edited by E. L. Berger (World Scientific, Singapore, 1992)}
\def \tasi90{{\it Testing the Standard Model} (Proceedings of the 1990
Theoretical Advanced Study Institute in Elementary Particle Physics,
Boulder, Colorado, 3--27 June, 1990), edited by M. Cveti\v{c} and P.
Langacker (World Scientific, Singapore, 1991)}

\def \zpc#1#2#3{Zeit. Phys. C {\bf#1}, #2 (#3)}

\begin{document}
\begin{titlepage}
{\large 	\hspace*{\fill} EFI 94-24\\
		\hspace*{\fill} hep-ph/9410267\\
		\hspace*{\fill} August 1994\\ }
\bigskip
\begin{center}
	{\Large \bf The heavy top quark in the two Higgs doublet model}\\
\end{center}
\bigskip
\begin{center}
	{\large Aaron K. Grant\\}
	Enrico Fermi Institute and Department of Physics\\
 	University of Chicago, Chicago, IL 60637\\
\end{center}
\medskip
\centerline{August 1994}
\bigskip
\begin{abstract}
Constraints on the
two Higgs doublet model are presented, assuming a top
mass of 174 $\pm$ 17 GeV.  We concentrate primarily on the ``type II''
model, where up--type quarks receive their mass from one Higgs doublet,
and down--type quarks receive their mass from the second doublet.  High energy
constraints derived from the $W$ mass, the full width of the $Z$
and the $b \bar b$ partial width of the $Z$ are combined with low
energy constraints from $\Gamma(b\rightarrow s \gamma)$,
$\Gamma(b \rightarrow c \tau \bar\nu_\tau)$ and $B^0$-$\bar B^0$ mixing
to determine the experimentally favored configurations
of the model.  This combination of observables rules out small charged Higgs
masses and small values of $\tan\beta$, and provides some information
about the neutral Higgs masses and the mixing angle $\alpha$.
In particular, constraints derived
from the $\rho$ parameter rule out configurations where the charged Higgs
is much heavier or much lighter than the neutral Higgses.  The agreement of the
model with experiment
is roughly as good as that found for the minimal standard model.
We discuss a scenario where
$\Gamma(Z\rightarrow b \bar b)$ is enhanced relative to the standard
model result, which unfortunately is on the verge of being ruled out by the
combination of $\Gamma(b\rightarrow s \gamma)$  and $\rho$
parameter constraints.  Implications for various extensions
of the standard model are briefly discussed.

\end{abstract}
\end{titlepage}

\renewcommand{\thesection}{\Roman{section}}
\renewcommand{\thetable}{\Roman{table}}

\section{Introduction}

Precise measurements of electroweak observables are beginning to place
significant constraints on extensions of the standard model.
Measurements of the $W$ mass \cite{Wmass}, the partial widths and
forward--backward asymmetries of the $Z$
\cite{LEP}, and the branching ratio for
$B\rightarrow X_s \gamma$ \cite{CLEO} all provide
significant information concerning unknown parameters of the standard
model as well as constraining
various types of new physics.  Furthermore, recent experimental evidence
for a top quark mass in the range $174\pm 17$ GeV \cite{CDFtop}
eliminates much of the uncertainty
in these constraints.  To date, the minimal standard model has
succeeded in its description of virtually all phenomena of electroweak origin
\cite{Rosner,Langacker}.  At the same time, however, there are some
sectors of the theory about which we know very little.  In
particular, predictions of the standard model depend very weakly
on the mass of the Higgs boson, and as a consequence we can say very
little about the Higgs sector of the theory.  In part for this reason,
the extension of the standard model to include a second Higgs doublet
is at present far from being ruled out experimentally.  Furthermore,
the inclusion of a second Higgs doublet is a common feature
of many extensions of the standard model, such as supersymmetry
or certain versions of the SO(10) and SU(5) grand unified theories.
It is therefore of interest to determine whether such a simple
extension of the standard model Higgs sector is compatible with experiment,
and what further extensions of the model are indicated if it is not.
In addition, we would like to determine whether the discovery of a charged
or pseudoscalar Higgs is necessarily an indication of supersymmetry, or whether
such a discovery could simply be an indication of an extended Higgs sector.

At tree level the two Higgs doublet model is identical, in most respects,
to the standard model.  It reproduces the important relation
$M_W^2=M_Z^2\cos^2\theta_W$, and furthermore the Yukawa couplings of
the quarks can be chosen so as to eliminate flavor changing neutral
Higgs interactions \cite{Glashow}.  The two Higgs model is also in
agreement with experiment at the level of radiative corrections.  The
two Higgs model differs from the standard model, however, in that
radiative corrections often depend rather sensitively on the details
of the Higgs sector.  For this reason, experimental data are beginning
to significantly constrain the parameters of the Higgs sector of the
two doublet model.  Low energy data such as $\Gamma(b \rightarrow s
\gamma)$, $\Gamma(b\rightarrow c \tau \bar\nu_\tau)$ and the $B^0-\bar
B^0$ mixing amplitude rule out small values of the charged Higgs mass,
and significantly constrain $\tan\beta$, the
ratio of the vacuum expectation values.
High energy data such as the
$W$ mass and the full width of the $Z$ constrain the Higgs masses and
mixing angles; in particular, configurations which give a positive
contribution to the $\rho$ parameter are disfavored by these data.
Additional stringent constraints are provided by $R_b\equiv
\Gamma(Z\rightarrow b\bar b)/\Gamma(Z\rightarrow{\rm hadrons})$.
Our aim here is to combine these constraints and to discuss the
implications of a moderately heavy top quark for the model.

In Sec.~II, we present a general overview of the two Higgs doublet model,
with particular emphasis on the structure of the Yukawa couplings.  In
Sec.~III, we discuss oblique corrections in the model, and discuss
their implications concerning the top quark mass.  In Sec.~IV, we
introduce constraints on the model from high energy data such as $R_b$
and the full $Z$ width.  As a byproduct, we demonstrate that the combination
of $\rho$ parameter constraints with constraints from $R_b$ makes the model
incompatible with a top mass larger than
about 200 GeV.  In Sec.~V, we introduce
constraints from low energy data such as $\Gamma(b\rightarrow s\gamma)$,
$\Gamma(b\rightarrow c \tau \bar\nu_\tau)$ and $B^0-\bar B^0$
mixing, and illustrate how these rule out small charged Higgs masses
and small values of $\tan\beta$.  In Sec.~VI
we combine all of the various constraints to determine the
configurations of the model that are compatible with experiment
for a top mass of 174 GeV.  In Sec.~VII we conclude.

\section{The two Higgs doublet model}

In this section we briefly review some generalities of the two Higgs
doublet model; for a more complete treatment, we refer the reader to
Refs.~\cite{HHG,Bertolini}.  The two Higgs doublet model is
essentially the $SU(2)\times U(1)$ standard model, modified by the
extension of the Higgs sector to include two scalar isodoublets
$\Phi_1,~\Phi_2$, which we write in component form as
\begin{equation}
\Phi_{i}=
\left( \begin{array}{c}  \phi_{i}^{+}\\ (\phi_{i}^{0}+ i \chi_{i}^{0})/\sqrt 2
 \end{array} \right).
\end{equation}
The Yukawa couplings of the scalars and fermions are constrained by
the requirement that there be no flavor changing neutral Higgs
interactions.  It was shown in Ref.~\cite{Glashow} that a necessary
and sufficient condition for the elimination of flavor changing
neutral interactions in the Higgs sector is that each quark of a given
charge must receive its mass from at most one Higgs field.  One finds
then that there are two possible arrangements for the Yukawa
couplings: one may either couple all of the fermions to a {\it single}
Higgs doublet (giving what is known as the ``type I'' model), or one
may couple $\Phi_1$ to the right--handed down--type quarks, and
$\Phi_2$ to the right--handed up--type quarks (giving what is known as
the ``type II'' model).  Here we will consider the latter possibility.
Imposing the discrete symmetry $\Phi_{2}\rightarrow -\Phi_{2}$,
$(u,c,t)_R\rightarrow -(u,c,t)_R$ is enough to ensure that the Yukawa
couplings have the correct form.  The Yukawa Lagrangian has the form
\begin{equation}
L_{Y}=\sum_{i=1}^{3} \bar\ell_{i,L}\Phi_1 e_{i,R} +\sum_{i,j=1}^{3}
D_{ij}\bar q_{i,L}\Phi_1 d_{j,R} +\sum_{i,j=1}^{3} U_{ij}\bar
q_{i,L}\Phi_2^c u_{j,R},
\end{equation}
where $\ell_{i,L},q_{i,L}$ are the left--handed lepton and quark
doublets, $U$ and $D$ are the quark mass matrices,
and $\Phi_2^c=-i\sigma_2\Phi_2^*$.
The Higgs potential, as in the standard model, has its
minimum at non-zero values of the fields $\Phi_1,~\Phi_2$.  After
applying the requirement that there be no flavor changing neutral
interactions, one finds that the vacuum expectation values (VEVs) can
be chosen both real and positive:
\begin{equation}
\langle\Phi_1 \rangle=
\left( \begin{array}{c}  0 \\ v_1/\sqrt 2
 \end{array} \right),~~~
\langle\Phi_2 \rangle=
\left( \begin{array}{c}  0 \\ v_2/\sqrt 2
 \end{array} \right).
\end{equation}
After diagonalizing the Higgs mass matrix, one finds that the fields
$\phi_{i}^{+}$ mix to form a charged Nambu--Goldstone boson $G^{+}$
and a charged physical scalar $H^{+}$ of mass $m_{+}$:
\begin{equation}
G^{+}=\cos\beta\phi_1^{+}+\sin\beta\phi_2^{+},~~~
H^{+}=-\sin\beta\phi_1^{+}+\cos\beta\phi_2^{+}.
\end{equation}
Similarly, the imaginary parts of the neutral components
$\chi^{0}_{1,2}$ mix to form a neutral Nambu--Goldstone boson $G^0$
and a $CP$--odd physical scalar $H_{3}^{0}$, again with mixing angle
$\beta$:
\begin{equation}
G^{0} = \cos\beta\chi_1^{0}+\sin\beta\chi_2^{0},~~~
H^{0}_{3}=-\sin\beta\chi_1^{0}+\cos\beta\chi_2^{0}.
\end{equation}
Finally, the scalars $\phi^{0}_{1,2}$ mix to form a pair of neutral
$CP$--even scalars, now with mixing angle $\alpha$:
\begin{equation}
H^{0}_{1}=\cos\alpha\phi_1^{0}+\sin\alpha\phi_2^{0},~~~
H^{0}_{2}=-\sin\alpha\phi_1^{0}+\cos\alpha\phi_2^{0}.
\end{equation}
The masses $m_{1,2}$ of $H^{0}_{1,2}$ obey $m_1\geq m_2$.

The mixing angle $\beta$ is given simply by
\begin{equation}
\tan\beta=\frac{v_2}{v_1}.
\end{equation}
The quantity $\tan\beta$ plays a central role in the theory because
the Yukawa couplings are often proportional to either $\tan\beta$ or
$\cot\beta$.  The Yukawa couplings of the various quarks to the
various scalars have been summarized in Table \ref{tab1}.
\begin{table}
\caption{Quark Yukawa couplings in the ``type II''
two doublet model.  Overall factors
of $-ig/2 M_W$, $-g/2 M_W$, $ig V_{ud} /M_W$ have been omitted in the
couplings of $H^0_{1,2}$, $H^0_3$, and $H^+$ respectively.}
\label{tab1}
\begin{center}
\begin{tabular}{cccc}
\hline\hline
Higgs & $u \bar u$ & $d \bar d$ & $\bar u d$ \\
\hline
$H_1^0$&$m_u\sin\alpha/\sin\beta$& $ m_d\cos\alpha/\cos\beta$ &-- \\
$H_2^0$&$m_u\cos\alpha/\sin\beta$& $-m_d\sin\alpha/\cos\beta$ &-- \\
$H_3^0$&$m_u\cot\beta\gamma_5$ &$m_d\tan\beta\gamma_5$ &-- \\ $H^+$ &
-- & -- & $m_u\cot\beta\bigl{(}\frac{1-\gamma_5}{2\sqrt{2}}\bigr{)}
+m_d\tan\beta\bigl{(}\frac{1+\gamma_5}{2\sqrt{2}}\bigr{)}$ \\
\hline\hline
\end{tabular}
\end{center}
\end{table}
We see in particular that couplings of up--type quarks are enhanced
for small values of $\tan\beta$, while couplings of down--type quarks
are enhanced for large values of $\tan\beta$.  Furthermore, one can
decouple $H^0_1$ from the up--type quarks by choosing $\alpha=0$, and
so on.  The large top quark mass, in combination with the enhancement
of the top Yukawa coupling for small values of $\tan\beta$ makes it
possible to derive a lower bound on $\tan\beta$.  Furthermore, we note
that since up--type quark masses are proportional to $v_2$ and
down--type masses to $v_1$, the mass hierarchy $m_t\gg m_b$ tends to
favor large $\tan \beta$.  In the type~I model, where all of the
quarks receive their masses from (say) $\Phi_1$, the Yukawa couplings
of the fermions are all proportional to $\cot\beta$ or $\csc\beta$;
consequently the Yukawa couplings of the bottom quark are negligible
compared to those of top quark regardless of the value of $\tan\beta$.

\section{Oblique Corrections in the Two Higgs Doublet Model}

In this section we discuss some features of oblique corrections to
electroweak observables in the two Higgs doublet model.

It has been known for quite some time that oblique radiative
corrections from the Higgs sector can be significantly larger in the
two Higgs model than in the minimal standard model
\cite{Bertolini,Toussaint,DGK}.  In particular, the contribution of
the Higgs sector to $\Delta\rho$, defined by
\begin{equation}
\Delta\rho=\frac{\Sigma_{WW}(0)}{M_W^2}-\frac{\Sigma_{ZZ}(0)}{M_Z^2},
\end{equation}
where $\Sigma_{WW,ZZ}$ are the $W$ and $Z$ self--energies, can be
significantly larger in the two doublet model than in the standard
model.  The non-standard contributions to the vector boson
self energies in the two doublet model
have been presented in Refs.~\cite{Bertolini, Hollik2}, and are
summarized in the Appendix.
In the minimal standard model, the contribution of the Higgs
to $\Delta\rho$ is given, in the limit $m_{\rm Higgs}\rightarrow\infty$, by
\begin{equation}
\Delta\rho_{\rm Higgs}(MSM)\simeq - \frac{3 G_F M_W^2}{8 \sqrt{2} \pi^2}
                \tan^2\theta_W\biggl{[}\log\frac{m_{\rm
Higgs}^2}{M_W^2}- \frac{5}{6} \biggr{]}
\end{equation}
and grows only logarithmically with the Higgs mass.  By contrast, the
top quark contribution to $\Delta\rho$ is given by
\begin{equation}
\Delta\rho_{\rm top}\simeq\frac{3 G_F}{8 \sqrt{2} \pi^2}m_t^2,
\end{equation}
which grows quadratically with $m_t$.  It is the weak dependence of
$\Delta\rho$ on $m_{\rm Higgs}$ that excludes top quark masses larger
than 200 GeV in the minimal standard model.  Since
$\Delta\rho$ grows quadratically with $m_t$, but falls off only logarithmically
with $m_{\rm Higgs}$, one must have an exponentially large Higgs mass
to prevent $\Delta\rho$ from becoming too large when the top is heavy;
since we expect on general grounds that
$m_{\rm Higgs}\leq 1~{\rm TeV}$, the standard model cannot accommodate
a top mass larger than about $200~{\rm GeV}$.  In the two Higgs doublet model,
the situation is different.  Here the leading behavior (for Higgs
masses much larger than the $W$ mass) of $\Delta\rho$ is given by
\cite{DGK}
\begin{equation}
\Delta\rho_{\rm Higgs}=\frac{3 G_F}{8 \sqrt{2} \pi^2}\biggl{(}
        \sin^2(\alpha-\beta) F(m_+,m_3,m_1)+\cos^2(\alpha-\beta)
F(m_+,m_3,m_2)\biggr{)},
\end{equation}
where
\begin{eqnarray}
F(m_+,m_3,m_{1,2})&=&m_+^2 -\frac{m_+^2
m_3^2}{m_+^2-m_3^2}\log\biggl(\frac{m_+^2}{m_3^2}\biggr)
\nonumber\\
&-&\frac{m_+^2 m_{1,2}^2}{m_+^2-m_{1,2}^2}
\log\biggl(\frac{m_+^2}{m_{1,2}^2}\biggr) +\frac{m_3^2
m_{1,2}^2}{m_3^2-m_{1,2}^2} \log\biggl(\frac{m_3^2}{m_{1,2}^2}\biggr).
\end{eqnarray}
The Higgs contribution to $\Delta\rho$ can be either positive or
negative depending on the ordering of the Higgs masses, and in general
grows quadratically with the largest Higgs mass. If the
Higgs masses are ordered as
\begin{equation}
\label{order}
m_{1,2}< m_{+}< m_3~~~~~{\rm or}~~~~~m_3< m_{+}< m_{1,2},
\end{equation}
then $\Delta\rho_{\rm Higgs}$ will be negative and grow quadratically
with the largest Higgs mass; if the charged Higgs is heavier or lighter than
all of the neutral Higgses, then $\Delta\rho_{\rm Higgs}$ will be positive.
The negative contribution to $\Delta\rho$ is largest if
$m_+\simeq 0.562 m_{\rm heavy}$, where $m_{\rm heavy}$ is the largest
Higgs mass.  For appropriate values of $\alpha-\beta$, one then has
\begin{equation}
\Delta\rho_{\rm Higgs}\simeq-\frac{3 G_F}{8 \sqrt{2} \pi^2}\times
0.216 m_{\rm heavy}^2.
\end{equation}
Since the top quark contribution to $\Delta \rho$ is quite large, the
Higgs contribution to $\Delta\rho$ must be either small (if it is positive)
or negative.

\section{Constraints from the full $Z$ width and $R_b$}

In this section we introduce constraints derived from the full $Z$
width and the $b\bar b$ partial width.  The LEP
measurements of these quantities \cite{LEP} are
\begin{equation}
\Gamma(Z\rightarrow {\rm all})= 2.4974 \pm 0.0038~{\rm GeV}
\end{equation}
and
\begin{equation}
R_b\equiv
\frac{\Gamma(Z\rightarrow b\bar b)}{\Gamma(Z\rightarrow{\rm hadrons})}=
0.2202\pm 0.0020.
\end{equation}
The quantity $R_b$ is particularly convenient to work with, since
oblique and QCD corrections to $\Gamma(Z\rightarrow b \bar b)$ and
$\Gamma(Z\rightarrow {\rm hadrons})$ cancel to a large extent in the
ratio $R_b$.  On the other hand, the full $Z$ width is rather sensitive
to oblique corrections through the $\rho$ parameter.  We begin with a
general discussion of the radiative corrections to
$\Gamma(Z\rightarrow f \bar f)$.

The $Z$ width in the two Higgs doublet model has been studied
previously in Ref.~\cite{DGHK}, and, in the context of the
supersymmetric standard model, in Ref.~\cite{Boulware}.  We have
independently carried out the calculation, and our results agree with
those found previously.  The vertex connecting the $Z$ to a pair of
fermions may be written as
\begin{equation}
\Gamma_{\mu}=i e \gamma_{\mu}(v-a \gamma_5),
\end{equation}
where at tree level the vector and axial couplings $v$ and $a$ are
\begin{equation}
v=\frac{I_3-2 Q s^2}{2 s c}~~~{\rm and}~~~a=\frac{I_3}{2 s c}.
\end{equation}
We use the abbreviations $s\equiv \sin \theta_W$, $c\equiv \cos
\theta_W$.  Radiative corrections modify these couplings through self
energy and vertex corrections.  The self energy corrections arise from
the $Z$ self energy and from the $Z$--photon mixing.  The $Z$ self
energy insertion introduces an overall factor of the $Z$ wavefunction
renormalization constant.  The $Z$--photon mixing modifies the vector
coupling of the $Z$ to a fermion pair: we make the replacement
\begin{equation}
v\rightarrow \frac{I_3-2 Q s^2}{2 s c} + Q
\frac{\Sigma_{AZ}(M_Z^2)}{M_Z^2}
\end{equation}
to include this effect.  Vertex diagrams introduce further corrections
to the vector and axial couplings of the Z, which we denote by $\delta
v$, $\delta a$.  Neglecting fermion masses, we find that
the partial $Z$ width into a particular final state
may then be written
\begin{equation}
\Gamma(Z\rightarrow f \bar f) = Z_Z
        \frac{n_c \alpha M_Z}{3}\biggl{(}v^2+a^2+2 v~{\rm Re}~\delta v
+ 2 a~{\rm Re}~\delta a + 2 v Q~{\rm
Re}~\frac{\Sigma_{AZ}(M_Z^2)}{M_Z^2} \biggr{)},
\end{equation}
where $n_c$ is the number of colors and $Z_Z$ is the $Z$ wavefunction
renormalization constant, which cancels in the ratio $R_b$.  We have employed
the renormalization scheme of Hollik \cite{Hollik} throughout our
calculations.

In addition to these purely electroweak corrections, partial widths
into quarks are modified by QCD corrections.  For quarks other than
the $b$, the QCD corrections enhance the partial width by about 4\%:
we have
\begin{equation}
\label{QCD}
\Gamma(Z\rightarrow q\bar q)=\Gamma^0(Z\rightarrow q\bar q)
	\biggl{[} 1+ \frac{\alpha_s(M_Z^2)}{\pi} + 1.41
\biggl{(}\frac{\alpha_s(M_Z^2)}{\pi}\biggr{)}^2\biggr{]},
\end{equation}
where $\Gamma^0$ is the uncorrected width.  In the case of $b$ quarks,
the finite $b$ mass and the large mass of the top quark modify the QCD
corrections.  The relevant corrections have been calculated in
Refs.~\cite{Kuhn1,Kuhn2,Kuhn3}.  In Ref.~\cite{Kuhn1}, it was pointed
out that the vector and axial parts of the $Z$ width into $b$ quarks
are affected differently by QCD. In particular, the vector--current
contribution to the $Z$ width into $b$ quarks is corrected by the same
QCD factor appearing in Eq.~(\ref{QCD}), while the axial part
$\Gamma_A$ is corrected by a factor
\begin{equation}
\Gamma_A=\Gamma_A^0 \biggl{[} 1+ \frac{\alpha_s(M_Z^2)}{\pi} +
        \biggr{(}1.41-I(z)/3\biggr{)}
\biggl{(}\frac{\alpha_s(M_Z^2)}{\pi}\biggr{)}^2\biggr{]},
\end{equation}
where $z=M_Z^2/4 m_t^2$, and $-I(z)/3$ is negative and increases in
magnitude for increasing top mass.  For a top mass of 100 GeV,
$-I(z)/3\sim -3$; for 250 GeV, $-I(z)/3\sim -5$.  The impact of these
QCD corrections is to slightly reduce $R_b$.  A second set of QCD
corrections has been discussed in Refs.~\cite{Kuhn2,Kuhn3}, where
corrections to $\Gamma(Z\rightarrow b\bar b)$ due to the finite $b$
mass have been calculated.  These corrections enhance
$\Gamma(Z\rightarrow b\bar b)$ by about 2 MeV, giving a small increase
in $R_b$.  The combined effect of all of these QCD corrections is to
enhance $R_b$ by about 0.4\%.

In Fig.~1 we have plotted $R_b$ as a function of the top mass for both
the minimal standard model and the two Higgs model for a few values of
$\tan\beta$.
\begin{figure}
\label{fig2}
\caption{$R_b$ as a function of the top mass in
the minimal standard model and in the two Higgs model.  The
solid curve is the standard model result for $m_{\rm Higgs}=100~{\rm
GeV}$.  The upper and lower dashed curves are the two Higgs results for
$m_2=m_3=50~{\rm GeV},~m_1=875~{\rm GeV},~m_+=422~{\rm GeV}$ and
$\tan\beta=70,~1$ respectively. For comparison, the standard model result
for $R_d$, the corresponding quantity for $d$ quarks, is also shown.  The
error bars indicate the $1\sigma$ experimental measurements of $m_t$ and
$R_b$. }
\begin{center}
        \BoxedEPSF{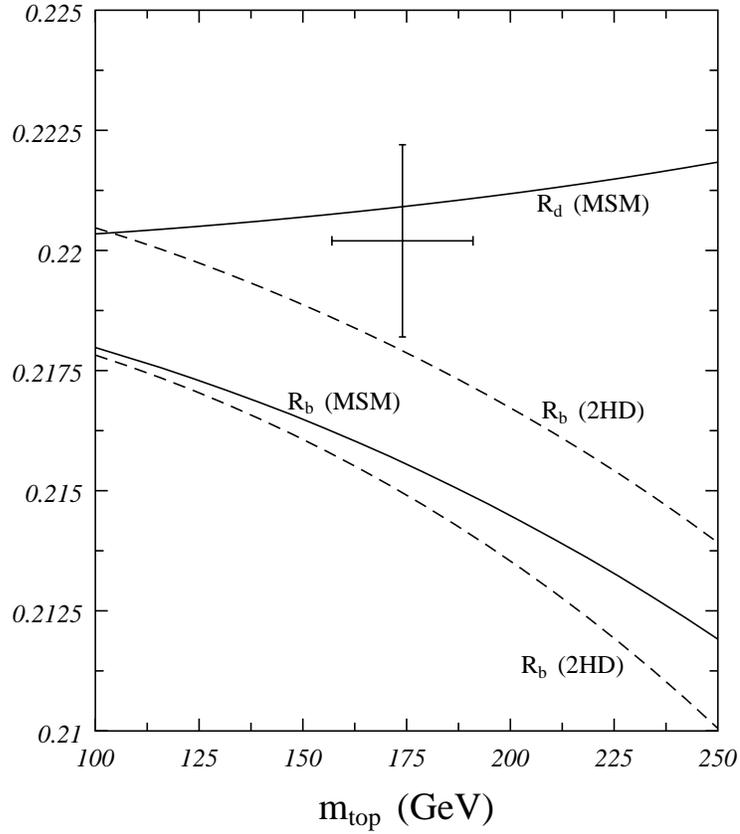 scaled 600}
\end{center}
\end{figure}

The $Z$ width into $b$ quarks is of particular interest because of the
presence of virtual top quarks in various vertex corrections.  These
corrections can be used to constrain $\tan\beta$ and the charged Higgs
mass, and, to a lesser extent, the mixing angle $\alpha$ and the
neutral Higgs masses.  The Higgs exchange vertex diagrams contributing
to $\Gamma(Z\rightarrow b \bar b)$ are shown in Fig.~2.
\begin{figure}
\label{fig3}
\caption{Vertex diagrams involving charged and neutral
        Higgs exchange in the two doublet model.  Diagrams involving
neutral Higgs exchange that do not contain potentially large
trigonometric factors (such as $\cot^2\beta,~\sec^2\beta$) have been
neglected.}
\begin{center}
        \BoxedEPSF{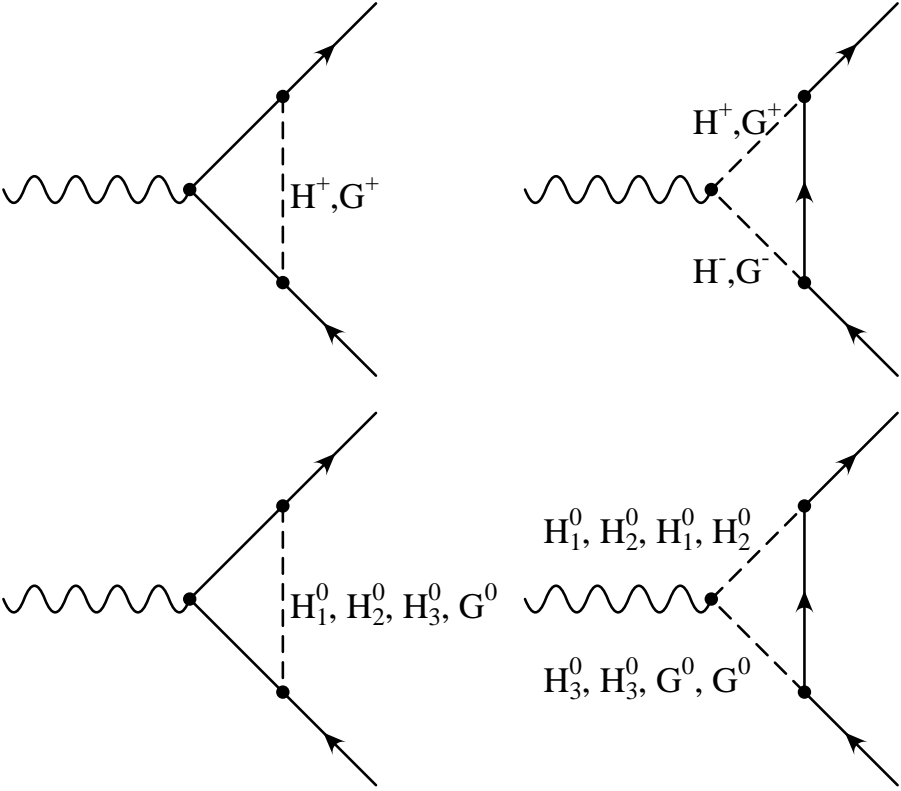 scaled 600}
\end{center}
\end{figure}
These break down into two classes: those involving charged Higgs
exchange, and those involving neutral Higgs exchange.  The dependence
of $R_b$ on the top mass results from the diagrams involving charged
Higgs exchange, since these involve the top quark Yukawa couplings of
Table~\ref{tab1}.  These diagrams tend to suppress the decay of the
$Z$ into $b$ quark pairs by an amount that grows quadratically with
the top mass.  In the minimal standard model, the effect of a heavy
top is to suppress the $Z$ width into $b$ quarks by an amount
\cite{Akhundov}
\begin{equation}
\Delta\Gamma(Z\rightarrow b \bar b)_{MSM}\simeq
        -\frac{\alpha^2_{QED} M_Z}{8 \pi s^2}\frac{1}{4 s^2 c^2}
\biggl{(}1-\frac{2 s^2}{3}\biggr{)} \biggl{[}\frac{m_t^2}{M_W^2}+
\biggl{(}\frac{8}{3}+\frac{1}{6 c^2}\biggr{)}\log\frac{m_t^2}{M_W^2}
\biggr{]}
\end{equation}
in the limit $m_t\gg M_W$.  In the two Higgs model, diagrams involving
charged Higgs exchange further suppress the $Z$ width into $b$ quarks
by an amount (for $m_+,~m_t\gg M_W$)
\begin{equation}
\Delta\Gamma(Z\rightarrow b \bar b)_{H^{\pm}}\simeq
        -\frac{\alpha^2_{QED} M_Z}{8 \pi s^2}\frac{1}{4 s^2 c^2}
\frac{m_t^2}{M_W^2\tan^2\beta} \biggl{(}1-\frac{2 s^2}{3}\biggr{)}
\biggr{(} -\frac{x}{(x-1)^2}\log x + \frac{x}{x-1} \biggr{)},
\end{equation}
where $x=m_t^2/m_+^2$.  This correction falls off for large $m_+$ and
grows with increasing $m_t$.  The corrections due to charged Higgs
exchange are always negative, and increase with increasing top mass.
This makes it difficult to accommodate a 150--200 GeV top quark mass,
particularly for small values of $\tan\beta$; for a top quark mass of
174 GeV, these diagrams exclude small values of $\tan\beta$.

The situation is somewhat different in the case of diagrams involving
neutral Higgs exchange.  Here we find that the contribution to
$\Gamma(Z\rightarrow b \bar b)$ can be either positive or negative,
depending on the masses of the neutral scalars.  These diagrams have
no top mass dependence, of course, since they involve only
$b$--quarks. Their explicit forms have been reported in Ref.~\cite{DGHK},
and are summarized in the Appendix.
These diagrams are only important for large values of
$\tan\beta$, for it is in this situation that the bottom quark's
Yukawa coupling becomes large.
In Fig.~3 we have plotted the
correction to $\Gamma(Z\rightarrow b \bar b)$ resulting from neutral
Higgs exchange as a function of $m_3$ for the special case $\alpha=\pi/2$,
$m_2=50~{\rm GeV}$, and $\tan\beta=70$.  In this case $H_1^0$
decouples from the vertex.  The correction is positive for small and
roughly equal masses $m_2\simeq m_3$, but becomes negative for large
mass splittings.  This is in fact one of the few situations where
vertex corrections significantly enhance the $Z$ partial width into
$b$ quarks.  These positive corrections make it possible to accommodate
a heavy top quark, particularly if the neutral Higgs bosons are light
and $\tan\beta$ is large.  As we will show below, however, large top quark
masses are not admissible in the model.

We have also studied the impact of these vertex corrections on
the partial widths for $Z\rightarrow \tau^+ \tau^-$ and
$Z\rightarrow\nu_\tau \bar\nu_\tau$.  Here the Higgs exchange vertex
corrections are important only
for large values of $\tan\beta$ and when two
or more  neutral scalars are light.  Both of the partial widths are
enhanced by a small amount by these corrections; however, the effect
is less than 1 MeV in both cases.

In light of the present discrepancy between the standard model prediction
of $R_b$ and the measured value, one interesting feature of the two
doublet model is the enhancement of $R_b$ relative to the standard model result
when $\tan\beta$ is large and two or more of the neutral scalars are light.
Although we will show below that such configurations of
the model are on the verge of being ruled out
by $\Gamma(b\rightarrow s\gamma)$, we
feel that direct experimental tests of this scenario are worth briefly
discussing.
Since the enhancement of $R_b$ requires small masses
for the neutral scalars,  it may be rather easy to test
at LEP II.  The cross section for the process
$e^+ e^- \rightarrow Z^* H^0_2$ is suppressed relative to the standard
model result for $e^+ e^- \rightarrow Z^* H^0$ by a factor of
$\sin^2(\alpha-\beta)$.  If $\sin^2(\alpha-\beta)$ is not too small, it should
be possible to detect a light scalar in the mass range
$m_2\sim 50-100~{\rm GeV}$ at LEP II using data from a standard model Higgs
search.  Barring extremely small values
of $\sin^2(\alpha-\beta)$, this should provide a fairly conclusive experimental
test of this scenario.  In the event that $\sin^2(\alpha-\beta)$
does turn out to be small, the process
$e^+ e^- \rightarrow Z^* \rightarrow H^0_2 H^0_3$, for which the cross section
is proportional to $\cos^2(\alpha-\beta)$, may provide a complementary
experimental probe of this scenario.
The current lower bound on $m_2$ \cite{ALEPH} ranges
from about 58.4 GeV for $\sin^2(\alpha-\beta)=1$ to about 30 GeV
for $\sin^2(\alpha-\beta)=0.05$.  Finally, we note that this scenario
is relevant only to the type II model.  In a type I model, the Yukawa
couplings of both top and bottom quarks are proportional to $\cot\beta$.
Consequently the positive vertex corrections from neutral Higgs
exchange are far smaller than the negative corrections
from charged Higgs exchange, and so one does not
expect significant enhancements of $\Gamma(Z\rightarrow b \bar b)$
at large values of $\tan\beta$.

\begin{figure}
\label{fig5}
\caption{       Correction to $\Gamma(Z\rightarrow b \bar b)$
                from $H^{0}_{2,3}$ exchange as a function of $m_3$
                for $m_2=50~{\rm GeV}$, $\tan\beta=70$, and $\alpha=\pi/2$.
	}
\begin{center}
        \BoxedEPSF{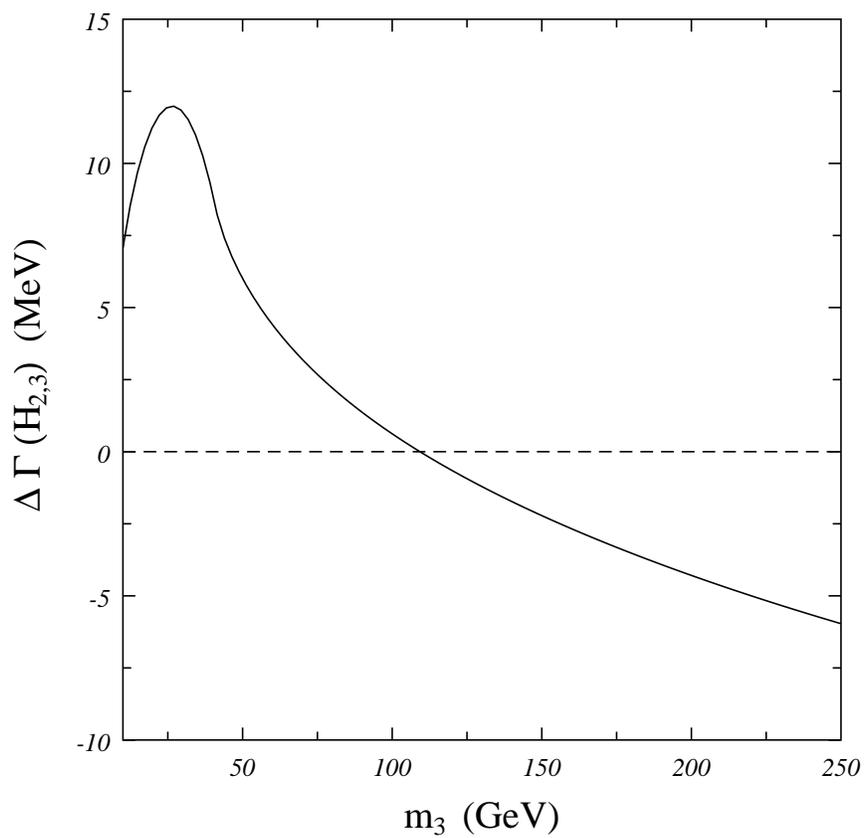 scaled 600}
\end{center}
\end{figure}

The various corrections to the $Z b \bar b$ vertex also modify the
$b \bar b$ forward--backward  asymmetry of the $Z$.  At tree level,
$A_{fb}(b \bar b)$ is given by
\begin{equation}
A_{fb}(b\bar b) = \frac{3}{4} \frac{2 v_e a_e}{v_e^2+a_e^2}
		\frac{2 v_b a_b}{v_b^2+a_b^2},
\end{equation}
where $v_f$, $a_f$ are the vector and axial couplings of the electron
and $b$ quark.
Vertex corrections tend to modify
the left--handed coupling of the $b$ for small values of $\tan\beta$, and
the right--handed coupling of the $b$ for large values of $\tan\beta$.
We have calculated the $b \bar b$ forward--backward asymmetry of the $Z$
including oblique and vertex corrections;  additional contributions
from box diagrams are known to be numerically small and so are neglected here.
The forward--backward asymmetry shows much the same behavior as $R_b$ as a
function of the Higgs masses and mixing angles:  $A_{fb}(b\bar b)$ is
suppressed at small values of $\tan\beta$ as a result of charged Higgs effects,
while in the case of large $\tan\beta$ there are several possibilities.
If all of the neutral Higgses are light (with masses less than 100 GeV)
and $\tan\beta$ is large, then $A_{fb}$ is enhanced slightly;  if
$H^0_2,H^0_3$ are light, but $H^0_1$ is much heavier, then $A_{fb}$ is
enhanced for $|\alpha|\simeq \pi/2$, but suppressed for $\alpha\simeq 0$.
In addition to its sensitivity to vertex corrections, $A_{fb}$ is
also sensitively dependent on oblique corrections through the value
of $\sin^2\theta_W$.  The enhancement of $A_{fb}$ at large values of
$\tan\beta$ limits the extent to which $R_b$ can be enhanced by neutral Higgs
effects.

To illustrate the range of top masses that can be accommodated by the
two doublet model, we can attempt to construct a scenario which allows
a very large top mass by arranging for negative contributions to
$\Delta\rho$
and positive contributions to $R_b$ from the Higgs sector.
In order to have  positive contributions to $R_b$, $m_3$ must be small.
The requirement of a small mass for $H^0_3$ together with the
fact that $m_2 \leq m_1$ implies that we must
have the mass hierarchy $m_2,m_3<m_+<m_1$.  We will then have a negative
contribution to $\Delta \rho$ for $|\alpha-\beta|=\pi/2$.
Here, however, we run
into a conflict with the the constraint from $R_b$:  in order to have
simultaneously $\tan\beta$ large and $|\alpha-\beta|=\pi/2$, we must
have $\beta\simeq \pi/2$, and $\alpha\simeq 0$.  Referring to
Table~\ref{tab1}, we see that for this value of $\alpha$
the Yukawa coupling of $H^0_2$ is small;  hence it is not
possible to  simultaneously obtain a negative contribution to the
$\rho$ parameter and still
have positive vertex corrections to $R_b$.
Although either constraint separately
can permit a very large top mass (on the order of 250 GeV or larger),
when the two are combined
the top mass is constrained to be less than about $200~{\rm GeV}$,
consistent
with both standard model expectations and recent evidence
for top quark production
at CDF \cite{CDFtop}.  The existence of this upper bound has been noted
previously by the authors of Ref.~\cite{GK}.
The mass limit of 200 GeV has been established
by a systematic search of the entire parameter space of the two doublet model.

The full $Z$ width provides constraints on the
charged Higgs mass.
The $Z$ width is quite sensitive to oblique
corrections through the $\rho$ parameter:  if the
$Z$ width is expressed in terms of $G_F$, the effect of oblique corrections
is to shift the value of $\sin^2 \theta_W$
and to renormalize the width by a factor $\rho$.  As a result,
positive contributions to $\Delta\rho$ tend to enhance the $Z$ width.
This constrains the mass
splittings in Higgs sector of the two doublet model;  in particular,
at the $1\sigma$ level and with a 174 GeV top quark,
we find that the charged Higgs mass is constrained by
\begin{equation}
m_{\rm light}- 130~{\rm GeV} \leq m_+ \leq m_{\rm heavy} +130~{\rm GeV},
\end{equation}
where $m_{\rm heavy, light}$ are the masses of the heaviest and lightest
neutral Higgses, respectively.  These mass limits have again been derived
by a systematic search of the entire parameter space of the two doublet model,
allowing each of the Higgs masses to vary between 50 and 1000 GeV, and each
of the mixing angles to vary over their full range.

\section{Constraints from $\Gamma(b \rightarrow s \gamma)$,
$B^0- \bar B^0$ mixing, and $\Gamma(b\rightarrow c \tau \bar\nu_\tau)$}

The CLEO collaboration \cite{CLEO}
has recently reported a value for the inclusive
branching ratio for radiative $B$ decays:
${\rm BR}(\bar B \rightarrow X_s \gamma)
= (2.32\pm 0.51 \pm 0.29 \pm 0.32)\times 10^{-4}$, where the first error
is statistical, and the latter two arise from systematic errors in the
yield and efficiency, respectively.  The 95\% confidence level limits
on the branching ratio are
\cite{CLEO}
\begin{equation}
1 \times 10^{-4} < {\rm BR}(\bar B \rightarrow X_s \gamma) < 4 \times 10^{-4}.
\end{equation}
This result can be used to rule out small charged Higgs masses and small values
of $\tan\beta$.
The relevant electroweak diagrams are shown in Fig.~4, and
have been calculated in Ref.~\cite{Hou}.
Ignoring QCD corrections,  the effective Lagrangian giving rise
to the $b \rightarrow s \gamma$ transition is
\begin{equation}
L_{\rm eff} = -\frac{G_F}{\sqrt{2}} V_{tb} V_{ts}^{*} \frac{e}{8 \pi^2} m_b
		\bar{s}_{L,\alpha} \sigma_{\mu \nu} b_L^{\alpha} F^{\mu\nu}
		\biggl{[} F(\frac{m_t^2}{M_W^2}) + G(\frac{m_t^2}{m_+^2})
		\biggr{]},
\end{equation}
where $\alpha$ is a color index, and $F$, $G$ are given by \cite{Hou}
\begin{eqnarray}
F(x)&=& \frac{3 x^3 -2 x^2}{4 (x-1)^4} \log x + \frac{-8 x^3 - 5 x^2 + 7 x}
		{24 (x-1)^3}\nonumber\\
G(y)&=& \frac{3 y^2 - 2 y}{6 (y-1)^3} \log y + \frac{-5 y^2 + 3 y}
		{12 (y-1)^2} \nonumber\\
		&+& \cot^2\beta \biggl{[}
		\frac{3 y^3-2 y^2}{12 (y-1)^4}\log y +
		\frac{ -8 y^3 - 5 y^2 +7 y}{72 (y-1)^3} \biggr{]}.
\end{eqnarray}
These results are only valid when one ignores QCD effects.  In order to
incorporate QCD, it is necessary to resum the large logarithms
$\sim \log(M_W^2/m_b^2)$ using the renormalization group.
The leading logarithmic QCD corrections result in an overall enhancement of
the $b \rightarrow s \gamma$ transition rate by a factor of 3 to 5 over
that which would be obtained using this na\"{\i}ve effective Lagrangian.
These corrections have been discussed extensively in the literature
\cite{Wise, Misiak, Ciuchini, Adel, Buras}.  Here we will use the QCD
calculations of Ref.~\cite{Ciuchini}. There are minor differences between
the various calculations of the anomalous dimension matrix; however
these result in only small (on the order of 1\% or less) variations
in the calculated value of ${\rm BR}( b\rightarrow s\gamma)$.  To extract
a value of ${\rm BR}(\bar B \rightarrow X_s \gamma)$, it is convenient
to use the approximation \cite{Wise}
\begin{equation}
\frac{\Gamma(\bar B \rightarrow X_s \gamma)}
     {\Gamma(\bar B \rightarrow X_c e \bar\nu_e)}\simeq
\frac{\Gamma( b \rightarrow s \gamma)}{\Gamma(b \rightarrow c e \bar\nu_e)},
\end{equation}
which reduces uncertainties due to a factor $m_b^5$ which appears
in both numerator and denominator.  Including the leading logarithmic QCD
corrections, we then have
\begin{equation}
\frac{\Gamma(\bar B \rightarrow X_s \gamma)}
     {\Gamma(\bar B \rightarrow X e \bar\nu_e)}\simeq
     \frac{6 \alpha}{\pi f(m_c/m_b)} \frac{|V_{ts}^* V_{tb}|^2}{|V_{bc}|^2}
     |A(m_b)|^2,
\end{equation}
where $A(m_b)$ is the coefficient of the operator $(e/8 \pi^2)
m_b\bar{s}_{L} \sigma_{\mu \nu} b_L F^{\mu\nu}$ evaluated at the scale
$m_b$ and $f(m_c/m_b)\simeq 0.316$ is the phase space factor for the
semi-leptonic decay.  The coefficient $A(m_b)$ is given in terms of
$\eta\equiv \alpha_s(m_b)/\alpha_s(M_W)$ by
\begin{equation}
A(m_b)=\eta^{-\frac{16}{23}}A(M_W)+ \frac{8}{3}(\eta^{-\frac{14}{23}}
       -\eta^{\frac{16}{23}}) B(M_W) + \sum_{i=1}^{8} \eta^{\alpha_i} x_i,
\end{equation}
where $A(M_W)=F(x)+G(y)$, $B(M_W)$ is the coefficient of the
``chromo--magnetic moment operator''
$(g/8 \pi^2) m_b\bar{s}^{\alpha}_{L}
\sigma_{\mu \nu} T^a_{\alpha\beta} b_L^{\beta} G^{\mu\nu}_a$ evaluated at the
scale $M_W$, and the sum
results from mixing with the four quark operator
$(\bar{c}_{L\alpha}\gamma^{\mu}b_{L\alpha})
(\bar{s}_{L\beta}\gamma_{\mu}c_{L\beta})$.  The numbers $\alpha_i$ and $x_i$
are related in a simple way to the eigenvalues and eigenvectors of the
anomalous dimension matrix.
We have used ${|V_{ts}^* V_{tb}|^2}/{|V_{bc}|^2}=0.95$.
\begin{figure}
\label{fig7}
\caption{Diagrams in the two Higgs model contributing to the magnetic moment
	 type operator that mediates the $b\rightarrow s \gamma$ transition.}
\begin{center}
        \BoxedEPSF{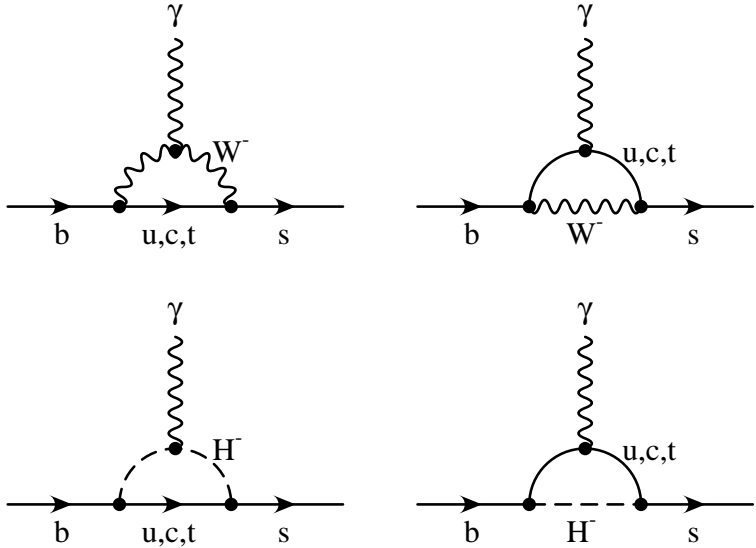 scaled 600}
\end{center}
\end{figure}

Recent analyses of the uncertainties in the leading logarithmic
calculation
have been given in Refs.~\cite{Buras,Ciuchini2}, where it was pointed out that
the theoretical uncertainty is dominated by the unknown next--to--leading
logarithmic corrections.   Other significant
uncertainties result from the experimental
error in the measurements of $\alpha_s(M_Z)=0.12\pm 0.01$ and the ratio
$m_c/m_b$, which
occurs in the phase space  factor for the semi--leptonic decay.
Combining the various theoretical
uncertainties given in Ref.~\cite{Buras} in quadrature, one finds that
the overall error for the standard model result
is on the order of 30\%; the error estimate
of Ref.~\cite{Ciuchini2} is also roughly of this magnitude. In the case
of the two doublet model, the situation is complicated by the greater
sensitivity of the result to the next--to--leading order corrections.
This has been pointed out in Ref.~\cite{Ciuchini2}.  In light of this,
the constraints from $b\rightarrow s \gamma$ presented here should
be considered as qualitatively correct, but nonetheless subject
to considerable theoretical uncertainty.  To factor in these uncertainties, in
deriving limits on $\tan\beta$ and the charged Higgs, we have
employed a 30\% estimate of the theoretical error. We require only
that
\begin{equation}
0.7\times ({\rm Theoretical~Estimate}) \leq {\rm Experimental~Upper~Bound}.
\end{equation}
This has the effect of
weakening somewhat the lower bound on the charged Higgs mass.

Additional low-energy constraints on the two doublet model can be obtained
using $B^0-\bar B^0$ mixing.  Here the relevant quantity is
$x_b\equiv\Delta m/\Gamma$,
where $\Delta m$ is the mass difference between the heavy and light
admixtures of $B^0$ and $\bar B^0$, and $\Gamma$ is their average width.
By observing time-dependent $B^0-\bar B^0$ oscillations, ALEPH and DELPHI
have determined the mass splitting to be $\Delta m=3.41\times10^{-4}$ eV,
which yields $x_b=0.797\pm0.128$ \cite{Venus}.
Measurements of the time integrated
mixing probability by ARGUS and CLEO give $x_b=0.67\pm0.08$ \cite{Venus}.
The weighted
average of these values is $x_b=0.706\pm0.068$.
In the two doublet model, $x_b$ is given by
\begin{equation}
x_b=\frac{\Delta m}{\Gamma} = \frac{G_F^2}{6 \pi ^2}|V_{td}^*|^2 |V_{tb}|^2
	f_B^2 m_B B_B \eta_B \tau_B M_W^2
	\biggl{(} I_{WW}+I_{WH}+I_{HH} \biggr{)},
\end{equation}
where $\tau_{B^0_d}=1.53\pm 0.09$ ps \cite{Venus}
is the $B$ lifetime, $f_B$ is the
$B$ meson decay constant, $m_B$ is the $B$ meson mass, $B_B$ is the
bag factor, and $\eta_B$ is a QCD correction factor, whose value lies
in the range from 0.55  to 0.85 \cite{Kim, BurasB, Schubert}.  The CKM matrix
element  $|V_{td}|$ lies in the range from 0.005 to 0.014 \cite{JonCKM},
while $V_{tb}=1$.
Finally, the dependence on $m_+$, $m_t$ and $\tan\beta$ is contained in
$I_{WW}$, $I_{WH}$, and $I_{HH}$.  We have \cite{HHG}
\begin{eqnarray}
I_{WW}&=& \frac{x}{4}\biggl{[}1+\frac{3-9x}{(x-1)^2}+
	\frac{6 x^2 \log x}{(x-1)^3}\biggr{]},\nonumber\\
I_{WH}&=&x y \cot^2\beta\biggr{[}\frac{(4 z-1)\log y}{2 (1-y)^2 (1-z)}
	-\frac{3 \log x}{2 (1-x)^2 (1-z)}
	+\frac{x-4}{(1-x)(1-y)} \biggr{]},~~{\rm and}\nonumber\\
I_{HH}&=&\frac{x y \cot^4 \beta}{4}\biggl{[} \frac{1+y}{(1-y)^2}
	+\frac{2 y \log y}{(1-y)^3}\biggr{]},
\end{eqnarray}
where $x=m_t^2/M_W^2$, $y=m_t^2/m_+^2$, and $z=M_W^2/m_+^2$.

A final constraint on the charged Higgs mass and $\tan\beta$ is provided
by $\Gamma(b\rightarrow c \tau \bar\nu_\tau)$.  Since in the two doublet
model this process is mediated by both $W$ and charged Higgs exchange, one
cannot have too small a mass for the charged Higgs or too large a value
of $\tan\beta$ without exceeding the experimental limits.  This
constraint has been analyzed recently in Ref.~\cite{bctau}, where it was
shown that the allowed values of $m_+$ and $\tan\beta$ must satisfy
\begin{equation}
\tan\beta\leq 0.51 m_+~[{\rm GeV}]
\end{equation}
at the $1\sigma$ level.
In contrast to the constraints given above, this measurement provides
an {\it upper} bound on $\tan\beta$, rather than a lower bound.

In Fig.~5 we have plotted the allowed region in the $m_+$-$\tan\beta$
plane after the constraints due to ${\rm BR}(b \rightarrow s \gamma)$,
$\Gamma(b\rightarrow c \tau \bar\nu_\tau)$, $B^0-\bar B^0$ mixing,
and $R_b$  have been applied, with a top mass of 174 GeV.  The constraints
from $R_b$ indicated are at the $2.5\sigma$
level; to agree with experiment to within
$2\sigma$ or less, $\tan\beta$ must be large
and the scalars $H^0_{2,3}$ must be light.
In this situation, the charged Higgs mass cannot be much larger than about
150 GeV, for reasons we will discuss in Sec.~VI.
We have used the 95\% confidence limits for $b\rightarrow s\gamma$.
The constraint from ${\rm BR}(b\rightarrow s \gamma)$  rules out
small charged Higgs masses, while the inclusion of $R_b$ strengthens the lower
bound on $\tan\beta$.  The constraint from $x_b$ is similar to that
from $R_b$ but somewhat weaker.
Similar conclusions have been reached in
Refs.~\cite{Buras,Ciuchini2,Park}.  From these considerations, we can constrain
$m_+$ and $\tan\beta$ by
\begin{equation}
m_+\geq 200~{\rm GeV}
\end{equation}
and
\begin{equation}
\tan\beta\geq 0.7
\end{equation}
at the $2.5\sigma$ level.  At the $2\sigma$ level, the bound
on $\tan\beta$ is roughly 13 for $m_t=157$~GeV, increasing
to 38 for $m_t=174$~GeV.

Since experimental
measurements of BR$(b\rightarrow s\gamma)$ are likely to improve,
we present also the charged Higgs mass limits
that result for smaller values of BR$(b\rightarrow s\gamma)$.  This
serves to illustrate the high sensitivity of these mass bounds to
theoretical uncertainties.  For comparison, we give both the mass
bound derived using the leading order QCD corrections we have employed
here, as well as the mass bounds reported in Ref.~\cite{Ciuchini2}
where the next--to--leading order corrections were partially included:
\begin{equation}
\label{charged_limits}
m_+\geq\cases{200~{\rm GeV~(LO)},~150~{\rm GeV~(NLO)}&
		BR$(b\rightarrow s \gamma)<4\times 10^{-4}$;\cr
\cr
	      380~{\rm GeV~(LO)},~200~{\rm GeV~(NLO)}&
		BR$(b\rightarrow s \gamma)<3\times 10^{-4}$;\cr
\cr
	      650~{\rm GeV~(LO)},~300~{\rm GeV~(NLO)}&
		BR$(b\rightarrow s \gamma)<2\times 10^{-4}$.\cr}
\end{equation}
We see that the lower bound on the charged Higgs
mass may move downward by a factor of two once the full next--to--leading
QCD corrections are known.
We see also that the partial inclusion of next--to--leading order
corrections relaxes the charged Higgs mass bound from the 200 GeV
cited above to about 150 GeV.

\begin{figure}
\label{fig8}
\caption{  Allowed region in the $m_+$--$\tan\beta$ plane for the combination
           of observables indicated.  The shaded region is allowed.
	 }
\begin{center}
        \BoxedEPSF{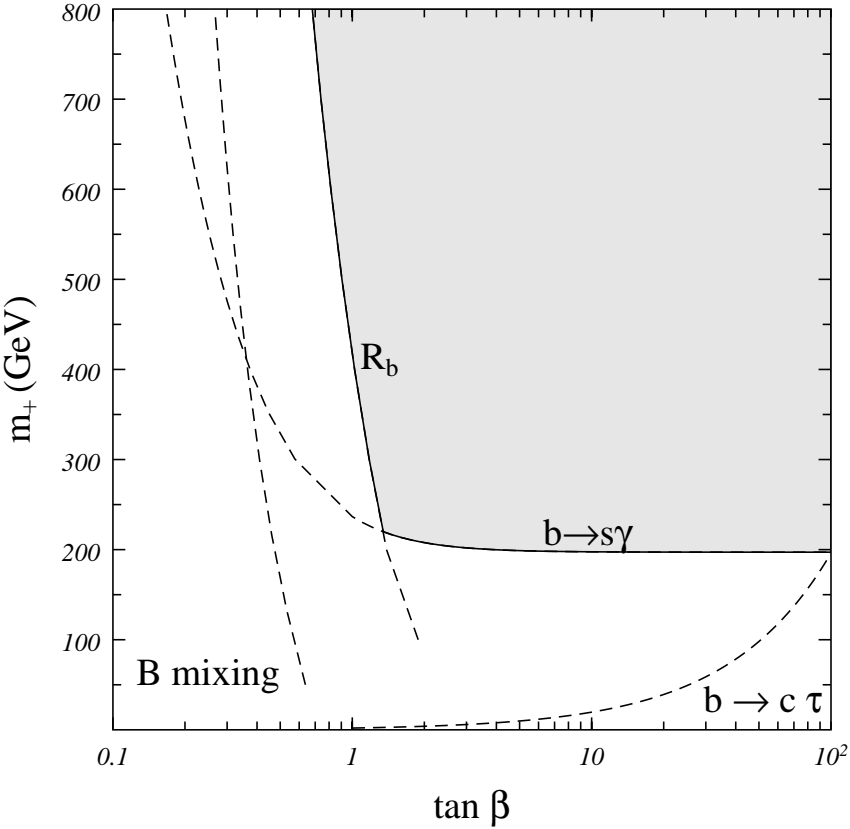 scaled 600}
\end{center}
\end{figure}

\section{Global constraints on the two doublet model}

In this section we discuss the experimentally preferred configurations
of the two doublet model by combining the constraints of the previous
sections. Throughout we assume a top quark mass of 174 GeV.
We first discuss those configurations
of the model which are compatible with each of
the various high energy measurements
($R_b$, $\Gamma(Z\rightarrow {\rm all})$, and $M_W$) at the $1.5\sigma$ level.
We then add to these the low energy constraints.

To be consistent with the LEP measurement of $R_b$ at the
$1.5\sigma$ level, one must assume that $\tan\beta$ is
large and that $H^0_2$ and
$H^0_3$ are quite light, having masses less than or of the order of 100 GeV.
We have evaluated the
maximum allowed masses and the minimum value of $\tan\beta$
by both a systematic search of the parameter
space and by randomly sampling a large number of
configurations of the model.
By both methods, we have found the limits
\begin{equation}
m_2,m_3\leq 65~{\rm GeV},
\end{equation}
and
\begin{equation}
\tan\beta\geq 53,
\end{equation}
which translates to a value of $\beta$ very nearly equal to $\pi/2$.
In addition, $R_b$ provides a constraint on the mixing angle $\alpha$ in the
case that $m_1$ is large; the reason is that, for $m_1$ large and
$\alpha\simeq 0$, the neutral Higgs contribution to $R_b$ is negative
(cf. Fig.~3).  As a consequence, one has
\begin{equation}
|\alpha|\geq \pi/3~~~~{\rm for}~~~~m_1\geq 100~{\rm GeV}.
\end{equation}

If one assumes the values of $m_{2,3}$ and $\tan\beta$ implied by $R_b$, then
further constraints can be derived from $M_W$ and the full
$Z$ width, through the $\rho$ parameter.
The leading behavior of $\Delta\rho$ in this case is given by
\begin{equation}
\label{deltarho}
\Delta\rho_{\rm Higgs} \simeq \frac{3 G_F}{8 \sqrt{2} \pi^2}
        \biggl{(}
	m_+^2-\cos^2\alpha\frac{m_+^2 m_1^2}{m_+^2-m_1^2}
        \log\biggl(\frac{m_+^2}{m_{1}^2}\biggr)
        \biggr{)}.
\end{equation}
{}From the leading behavior of $\Delta\rho_{\rm Higgs}$,
we see immediately
that the charged Higgs mass cannot be too large,
for this would give
a large positive contribution to $\Delta\rho$.  Furthermore, although the
second term in Eq.~(\ref{deltarho}) is negative, its coefficient
is small due to the bound on $\alpha$ imposed by $R_b$ when $m_1$ is large.
As a result, we have
\begin{equation}
\label{bounds}
m_+\leq 150~{\rm GeV}.
\end{equation}
Due to the quadratic growth of $\Delta\rho$ with $m_+$, larger values
are strongly disfavored.
Finally, large values of $m_1$ are {\it weakly} favored by $\rho$ parameter
constraints;  one has $m_1\geq 500$ GeV, although smaller values are
not in terribly poor agreement with the data.
In combination with constraints from $R_b$, the bound on the mass of $m_1$
gives
\begin{equation}
|\alpha|\geq\pi/3.
\end{equation}
This configuration of the model is also compatible with the $b \bar b$
forward-backward asymmetry of the $Z$.
In deriving these limits, we have again carried out a systematic search
of the parameter space of the model, covering the full range of $\alpha$
and $\beta$, and varying each of the scalar masses between 50 and 1000 GeV.
If one allows larger values of $m_1$, then slightly larger values of $m_+$ are
admissible; for $m_1=1.5$ TeV, the upper limit on the charged Higgs
mass recedes to 170 GeV.

Finally, we can combine the constraints from high energy data with
those from $b\rightarrow s \gamma$, $b\rightarrow c \tau \bar\nu_\tau$, and
$B$ mixing.
This class of constraints also rules out small values of $\tan\beta$,
although the bounds are weaker than those derived from $R_b$.
If we take the weaker of the two charged Higgs mass bounds given in
Eq.~(\ref{charged_limits}), we find that the charged Higgs mass
is constrained to lie in a narrow region in the vicinity of 150 GeV:
\begin{equation}
m_+ \simeq 150~{\rm GeV}.
\end{equation}
We find that compatibility
of the model with the measured value of $R_b$, when combined with the
other relevant observables, yields a very restrictive set of constraints.

\section{CONCLUSIONS}

We have presented constraints on the type II two Higgs doublet model from a
variety of observables.  The overall agreement of the model with
experiment is acceptable, and quite comparable to that found for the minimal
standard model.  We have shown that $\tan\beta$ and the charged Higgs
mass are bounded by
\begin{equation}
m_+\geq 150 \sim 200~{\rm GeV},~~~\tan\beta\geq 0.7
\end{equation}
for $2.5\sigma$ agreement with measurements of $R_b$, and discussed
the sensitivity of the charged Higgs mass bound to theoretical
uncertainties.  The lower bound on $\tan\beta$ increases to about
13 for $2\sigma$ agreement with $R_b$.
Oblique constraints on the model require that the
charged Higgs mass lie in the range
\begin{equation}
m_{\rm light}-130~{\rm GeV}\leq m_+\leq m_{\rm heavy} + 130~{\rm GeV},
\end{equation}
where $m_{\rm heavy, light}$ are the masses of the heaviest and lightest
neutral Higgses.
We have discussed one very tightly constrained
scenario where the model (unlike the standard model)
agrees with measurements of $R_b$ at the 1 to $1.5\sigma$ level.
In this case, every parameter of the model is constrained in some way;
for the mixing angles one has
\begin{equation}
\tan\beta\geq 53,~~~~|\alpha|\geq\pi/3,
\end{equation}
while for the Higgs masses one finds
\begin{equation}
m_{2,3}\leq 65~{\rm GeV},~~~m_+\simeq 150~{\rm GeV},~~~m_1\geq 500~{\rm GeV}.
\end{equation}
The preference for large values of $m_1$ is weak, and $\rho$ parameter
constraints strongly
disfavor larger values of $m_+$.  This configuration of the model is
virtually ruled out by $b\rightarrow s\gamma$: only a very narrow
allowed region remains for the charged Higgs mass, and if the uncertainty
on the measurement is reduced even slightly, this window is likely to
disappear.
Consequently, it is improbable
that the two doublet model can improve on the standard model prediction
of $R_b$.

The large $\tan\beta$ scenario discussed here may
nonetheless be of some interest
for studies of supersymmetric grand unified theories.  Indeed,
if one assumes a large value of $\tan\beta$ and a
charged Higgs mass on the order of 100 GeV, then one
finds that $H^0_{2,3}$ are light and nearly degenerate, and
that $\alpha$ is very nearly equal to $-\pi/2$.
This configuration of the model is surprisingly similar to that found
in the large $\tan\beta$ scenario.  In
supersymmetric SO(10) models the Yukawa couplings of the $t$, $b$, and
$\tau$ can be made to unify at the unification scale if $\tan\beta\simeq 50-60$
and $m_t\simeq 160-190$ GeV \cite{Polonsky, Hall}.  Of course, the full
calculation including superpartners is necessary
in order to study the viability of this scenario.

We have shown that the type II two Higgs doublet model is roughly
consistent with
numerous experimental observations at the level of radiative corrections.
The measured values of the $W$ mass, the $Z$ width, and low energy data
such as $\Gamma(b \rightarrow s\gamma)$,
$\Gamma(b\rightarrow c \tau\bar\nu_\tau)$ and the $B^0-\bar B^0$
mixing amplitude can all be accommodated within the
model.  The partial width of the $Z$ into $b$ quarks poses a significant
challenge to the model, which will become increasingly restrictive
as data on this and other observables improve.  We conclude, then,
that the minimal two doublet model without supersymmetry cannot be ruled
out on the basis of current data.  Although the discovery of a light
neutral Higgs or a charged Higgs would be indicative of
supersymmetry,  it is not an ironclad guarantee that the world is
indeed supersymmetric.

\section{Acknowledgments}

I would like to thank Tony Gherghetta,
Emmin Shung, and Mihir Worah
for enlightening discussions.  I am particularly indebted to Jonathan Rosner
for originally suggesting the topic of this work, for critically reading
the manuscript, and for numerous helpful suggestions. This work was
supported in part by the U. S. Department of Energy
under Grant No. DE~FG02~90ER~40560.

\newpage

\section{APPENDIX}

In this appendix we collect the vector boson self energies and
vertex corrections necessary for our calculations.  The vector boson self
energies can be expressed in terms of the two point integral
\begin{equation}
I_0(p^2,m_1,m_2)=\frac{16\pi^2}{i}\mu^{(4-D)}\int d^D k
	\frac{1}{(k^2-m_1^2)([k-p]^2-m_2^2)}.
\end{equation}
Another frequently appearing combination is
\begin{equation}
\Delta I_0(p^2,m_1,m_2)\equiv\frac{I_0(p^2,m_1,m_2)-I_0(0,m_1,m_2)}{p^2}.
\end{equation}
For the vertex corrections we need in addition the vector two point integral
$I_1$, defined by
\begin{equation}
p_\mu I_1(p^2,m_1,m_2)=\frac{16\pi^2}{i}\mu^{(4-D)}\int d^D k
	\frac{k_\mu}{(k^2-m_1^2)([k-p]^2-m_2^2)},
\end{equation}
the scalar three point integral $K_0$,
\begin{equation}
K_0=\frac{16\pi^2}{i}\int d^4 k \frac{1}{(k^2-m_1^2)
	([k-p_2]^2-m_2^2)([k+p_1]^2-m_3^2)},
\end{equation}
and the tensor three point integrals $K_{11,22,12,00}$ defined by
\begin{eqnarray}
p_{1\mu}p_{1\nu}K_{11}+p_{2\mu}p_{2\nu}K_{22}
+(p_{1\mu}p_{2\nu}+p_{1\nu}p_{2\mu})K_{12}+g_{\mu\nu}K_{00}&=&\nonumber\\
\frac{16\pi^2}{i}\mu^{(4-D)}\int d^D k \frac{k_\mu k_\nu}{(k^2-m_1^2)
	([k-p_2]^2-m_2^2)([k+p_1]^2-m_3^2)}&&.
\end{eqnarray}
The arguments of the three point integrals are
$(q^2,p_1^2,p_2^2,m_1,m_2,m_3)$, where $p_{1,2}$ are the
momenta of the final state fermions and $q=p_1+p_2$.
The arguments of these integrals are given below in the form
$(q^2,m_1,m_2,m_3)$.
The unrenormalized vector boson self energies
$\Sigma_{AA}$, $\Sigma_{AZ}$, $\Sigma_{ZZ}$, and $\Sigma_{WW}$ are
\begin{eqnarray}
\Sigma_{AA}(p^2)&=&\frac{\alpha}{4\pi}\biggl\{
		\sum_{\rm f} \frac{4 N_C^f Q_f^2}{3}
		\biggl[ (p^2+2 m_f^2)I_0(p^2,m_f,m_f)
		-2 m_f^2 I_0(0,m_f,m_f)-p^2/3\biggr]\nonumber\\
		&-&\biggl[ (3p^2+4M_W^2)I_0(p^2,M_W,M_W)-4M_W^2I_0(0,M_W,M_W)
		 \biggr]\nonumber\\
		&-&\biggl[
			(4m_+^2-p^2)I_0(p^2,m_+,m_+)/3-4m_+^2I_0(0,m_+,m_+)/3
			-2p^2/9\biggr]\biggr\},
\end{eqnarray}
\begin{eqnarray}
\Sigma_{AZ}(p^2)&=&-\frac{\alpha}{4\pi}\biggl\{
		\sum_{\rm f} \frac{4 N_C^f v_f Q_f}{3}
		\biggl[ (p^2+2 m_f^2)I_0(p^2,m_f,m_f)
		-2 m_f^2 I_0(0,m_f,m_f)-p^2/3\biggr]\nonumber\\
		&-&\frac{1}{3sc}\biggl[
		\biggl([9c^2+1/2]p^2+[12c^2+4]M_W^2\biggr)I_0(p^2,M_W,M_W)
		\nonumber\\
		&-&(12 c^2-2)M_W^2I_0(0,M_W,M_W)+p^2/3\biggr]\nonumber\\
		&-&\frac{c^2-s^2}{3sc}\biggl[
		(2m_+^2-p^2/2)I_0(p^2,m_+,m_+)-2m_+^2I_0(0,m_+,m_+)
		-p^2/3\biggr]\biggr\},
\end{eqnarray}
\begin{eqnarray}
\Sigma_{ZZ}(p^2)&=&\frac{\alpha}{4\pi}\biggl\{
		\sum_{\rm f} \frac{4 N_C^f}{3}\biggl[
		\bigl(v_f^2+a_f^2\bigr)\biggl(
		[p^2+2 m_f^2]I_0(p^2,m_f,m_f)
		-2 m_f^2 I_0(0,m_f,m_f)-p^2/3\biggr)\nonumber\\
		&-&\frac{m_f^2}{2s^2c^2}I_0(p^2,m_f,m_f)\biggr]\nonumber\\
		&-&\frac{1}{6s^2c^2}\biggr[
		\bigg([24 c^4+16 c^2-10]M_W^2+[18c^4+2c^2-1/2]p^2\biggr)
		I_0(p^2,M_W,M_W)\nonumber\\
		&-&(24c^4-8c^2+2)M_W^2I_0(0,M_W,M_W)+(4c^2-1)p^2/3\biggr]
		\nonumber\\
		&-&\frac{1}{12s^2c^2}\biggl[
		(c^2-s^2)^2\biggl([4m_+-p^2]I_0(p^2,m_+,m_+)
		-4m_+^2 I_0(0,m_+,m_+)-2p^2/3\biggr)\nonumber\\
		&+&\cos^2(\alpha-\beta)\biggr(
		[2m_1^2-10M_Z^2-p^2]I_0(p^2,m_1,M_Z)-[m_1^2-M_Z^2]^2
		\Delta I_0(p^2,m_1,M_Z)\nonumber\\
		&-&2m_1^2I_0(0,m_1,m_1)-2M_Z^2I_0(0,M_Z,M_Z)-2p^2/3\biggl)
		\nonumber\\
		&+&\sin^2(\alpha-\beta)\biggr(
		[2m_2^2-10M_Z^2-p^2]I_0(p^2,m_2,M_Z)-[m_2^2-M_Z^2]^2
		\Delta I_0(p^2,m_2,M_Z)\nonumber\\
		&-&2m_2^2I_0(0,m_2,m_2)-2M_Z^2I_0(0,M_Z,M_Z)-2p^2/3\biggl)
		\nonumber\\
		&+&\sin^2(\alpha-\beta)\biggr(
		[2m_1^2+2m_3^2-p^2]I_0(p^2,m_1,m_3)-[m_1^2-m_3^2]^2
		\Delta I_0(p^2,m_1,m_3)\nonumber\\
		&-&2m_1^2I_0(0,m_1,m_1)-2m_3^2I_0(0,m_3,m_3)-2p^2/3\biggl)
		\nonumber\\
		&+&\cos^2(\alpha-\beta)\biggr(
		[2m_2^2+2m_3^2-p^2]I_0(p^2,m_2,m_3)-[m_2^2-m_3^2]^2
		\Delta I_0(p^2,m_2,m_3)\nonumber\\
		&-&2m_2^2I_0(0,m_2,m_2)-2m_3^2I_0(0,m_3,m_3)-2p^2/3\biggl)
		\biggr]
		\biggr\},
\end{eqnarray}
\begin{eqnarray}
\Sigma_{WW}(p^2)&=&\frac{\alpha}{4\pi}\biggl\{
		\sum_{\rm f} \frac{N_C^f}{6s^2}\biggl[
		(2p^2-m_f^2-m_{f'}^2)I_0(p^2,m_f,m_{f'})
		\nonumber\\
		&-&2m_f^2I_0(0,m_f,m_f)-2m_{f'}^2I_0(0,m_{f'},m_{f'})
		-(m_f^2-m_{f'}^2)^2 \Delta I_0(p^2,m_f,m_{f'})-2p^2/3\biggr]
		\nonumber\\
		&-&\frac{2}{3}\biggl[(5 p^2+2 M_W^2)I_0(p^2,M_W,\lambda)
		-2M_W^2I_0(0,M_W,M_W)-M_W^4\Delta I_0(p^2,M_W,\lambda)+p^2/3
		\biggr]
		\nonumber\\
		&-&\frac{1}{12s^2}\biggl[
		\biggl([40c^2-1]p^2+[16c^2+54-10/c^2]M_W^2\biggr)
		I_0(p^2,M_W,M_Z)
		\nonumber\\
		&-&(16c^2+2)M_W^2I_0(0,M_W,M_W)-(16c^2+2)M_Z^2I_0(0,M_Z,M_Z)
		\nonumber\\
		&+&(8c^2-2)p^2/3-(1+8c^2)(M_W^2-M_Z^2)^2
		\Delta I_0(p^2,M_W,M_Z)\biggr]
		\nonumber\\
		&-&\frac{1}{12s^2}\biggl[
		\cos^2(\alpha-\beta)\biggl([2m_1^2-10M_W^2-p^2]I_0(p^2,M_W,m_1)
		-[m_1^2-M_W^2]^2\Delta I_0(p^2,M_W,m_1)\nonumber\\
		&-&2m_1^2 I_0(0,m_1,m_1)-2M_W^2 I_0(0,M_W,M_W)-2p^2/3\biggr)
		\nonumber\\
		&+&\sin^2(\alpha-\beta)\biggl([2m_2^2-10M_W^2-p^2]
		I_0(p^2,M_W,m_2)
		-[m_2^2-M_W^2]^2\Delta I_0(p^2,M_W,m_2)\nonumber\\
		&-&2m_2^2 I_0(0,m_2,m_2)-2M_W^2 I_0(0,M_W,M_W)-2p^2/3\biggr)
		\nonumber\\
		&+&
		\sin^2(\alpha-\beta)\biggl([2m_1^2+2m_+^2-p^2]I_0(p^2,m_+,m_1)
		-[m_1^2-m_+^2]^2\Delta I_0(p^2,m_+,m_1)\nonumber\\
		&-&2m_1^2 I_0(0,m_1,m_1)-2m_+^2 I_0(0,m_+,m_+)-2p^2/3\biggr)
		\nonumber\\
		&+&\cos^2(\alpha-\beta)\biggl([2m_2^2+2m_+^2-p^2]
		I_0(p^2,m_+,m_2)
		-[m_2^2-m_+^2]^2\Delta I_0(p^2,m_+,m_2)\nonumber\\
		&-&2m_2^2 I_0(0,m_2,m_2)-2m_+^2 I_0(0,m_+,m_+)-2p^2/3\biggr)
		\nonumber\\
		&+&\biggl([2m_3^2+2m_+^2-p^2]
		I_0(p^2,m_+,m_3)
		-[m_3^2-m_+^2]^2\Delta I_0(p^2,m_+,m_3)\nonumber\\
		&-&2m_3^2 I_0(0,m_3,m_3)-2m_+^2 I_0(0,m_+,m_+)-2p^2/3\biggr)
		\biggr]
		\biggr\}.
\end{eqnarray}
The sum $\sum_{\rm f}$ is over all fermions in the theory, and $f'$ is the
isospin partner of $f$.  We have neglected quark mixing in the $W$ self energy.

In the calculation of the $Z$ width, we need in addition the various
vertex corrections.  Here we tabulate only those corrections involving
exchange of physical scalars.  For the full calculation, one must also
include diagrams involving $W$ and $Z$ exchange, which are present also
in the minimal standard model.  For the pure standard model corrections,
we refer the
reader to Refs.~\cite{DGHK,Hollik}.  Here we give the {\it renormalized}
vertex corrections to $Z\rightarrow b\bar b$ from physical Higgs exchange.
Where no ambiguity is involved, the arguments of groups of tensor integrals
that depend on the same variables have been specified only once at the end
of the group.

(a)  Charged Higgs vertex correction:

\begin{eqnarray}
\delta\Gamma_\mu &=& ie\gamma_\mu \frac{\alpha}{8 \pi s^2}
		\biggl\{
		\frac{m_t^2\cot^2\beta}{M_W^2}
		\biggl(\frac{1-\gamma_5}{2}\biggr)
		\biggl(
		\frac{s^2-c^2}{sc}K_{00}(M_Z^2,m_t,m_+,m_+)\nonumber\\
		&+&
		[v_t-a_t][2 K_{00}-1/2+M_Z^2 K_{12}-m_t^2K_0]
		(M_Z^2,m_+,m_t,m_t)-\frac{m_t^2}{2 s c} K_0(M_Z^2,m_+,m_t,m_t)
		\nonumber\\
		&-&[v_b+a_b]I_1(0,m_t,m_+)
		\biggr)\nonumber\\
		&+&\frac{m_b^2\tan^2\beta}{M_W^2}
		\biggl(\frac{1+\gamma_5}{2}\biggr)
		\biggl(\frac{s^2-c^2}{sc}K_{00}(M_Z^2,m_t,m_+,m_+)
		+[v_t+a_t][2 K_{00}-1/2\nonumber\\
		&+&M_Z^2K_{12}-m_t^2K_0]
		(M_Z^2,m_+,m_t,m_t)
		+\frac{m_t^2}{2sc}K_0(M_Z^2,m_t,m_+,m_+)
		\nonumber\\
		&-&[v_b-a_b]I_1(0,m_t,m_p)
		\biggr)\biggl\}
\end{eqnarray}

(b) Neutral Higgs vertex correction:

\begin{eqnarray}
\delta\Gamma_\mu &=&
		ie\gamma_\mu \frac{\alpha}{8 \pi s^2}\frac{m_b^2}{2 M_W^2}
		\sum_{\pm}\biggl\{
		\biggl(\frac{1\pm\gamma_5}{2}\biggr)
		\biggl[\pm\frac{2\sec\beta}{sc}\biggl(
		\tan\beta\cos\alpha\sin(\beta-\alpha)
		K_{00}(M_Z^2,0,m_1,m_3)\nonumber\\
		&+&\tan\beta\sin\alpha\cos(\beta-\alpha)
		K_{00}(M_Z^2,0,m_2,m_3)
		+\cos\alpha\cos(\beta-\alpha)
		K_{00}(M_Z^2,0,m_1,M_Z)\nonumber\\
		&-&\sin\alpha\sin(\beta-\alpha)
		K_{00}(M_Z^2,0,m_2,M_Z)\biggr)
		\nonumber\\
		&+&(v_b\pm a_b)\biggl(
		\frac{\cos^2\alpha}{\cos^2\beta}
		(2K_{00}-1/2+M_Z^2 K_{12})(M_Z^2,m_1,0,0)\nonumber\\
		&+&\frac{\sin^2\alpha}{\cos^2\beta}
		(2K_{00}-1/2+M_Z^2 K_{12})(M_Z^2,m_2,0,0)
		+\tan^2\beta
		(2K_{00}-1/2+M_Z^2 K_{12})(M_Z^2,m_3,0,0)\nonumber\\
		&+&(2K_{00}-1/2+M_Z^2 K_{12})(M_Z^2,m_Z,0,0)
		\biggr)
		-(v_b\mp a_b)\biggl(
		\frac{\cos^2\alpha}{\cos^2\beta}
		I_1(0,m_1,0)\nonumber\\
		&+&\frac{\sin^2\alpha}{\cos^2\beta}
		I_1(0,m_2,0)
		+\tan^2\beta
		I_1(0,m_3,0)
		+I_1(0,M_Z,0)
		\biggr)
		\biggr]\biggr\}
\end{eqnarray}

\end{document}